# Laser-assisted high-energy proton pulse extraction for feasibility study of co-located muon source at the SNS

Yun Liu[1], Abdurahim Rakhman, Cary D. Long

Research Accelerator Division, Oak Ridge National Laboratory, Oak Ridge National Laboratory, Oak Ridge, TN 37831, USA.

Yuan Liu

Physics Division, Oak Ridge National Laboratory, Oak Ridge National Laboratory, Oak Ridge, TN 37831, USA.

Travis J. Williams[2]

Neutron Scattering Division, Oak Ridge National Laboratory, Oak Ridge National Laboratory, Oak Ridge, TN 37831, USA.

[1] liuy2@ornl.gov
[2] williamstj@ornl.gov

**Abstract**

We have experimentally demonstrated the first non-intrusive 1-GeV proton beam extraction for the generation of muons with a temporal structure optimized for Muon Spin Relaxation/Rotation/Resonance (µSR) applications. The proton pulses are extracted based on the laser neutralization of 1 GeV hydrogen ion ($H^-$) beam in the high energy beam transport of the Spallation Neutron Source (SNS) accelerator. The maximum flux of the extracted proton beam accounts for only 0.2% of the total proton beam used for neutron production, a marked difference from the 20% reduction at other co-located muon and neutron facilities, and thus the proposed method will result in negligible impact on the SNS operation. This paper describes the development of a fiber/solid-state hybrid laser system that has high flexibility of pulse structure and output power, initial experiments on laser neutralization of $H^-$ beam and separation of $H_0$ beam from the existing SNS accelerator beam line, conversion of $H_0$ to proton at the SNS linac dump, and measurement results of 30 ns/50 kHz proton pulses. This system conclusively demonstrates the feasibility of laser-based proton beam extraction to power a world-leading µSR facility at the SNS.

1. **Introduction**

Muon Spin Rotation/Relaxation/Resonance (µSR) is an experimental technique that involves using spin-polarized muons implanted in a material to provide extremely sensitive measurements of the static and dynamic properties of the local magnetic field distribution within the sample. This mature technique has been used to provide important scientific measurements in condensed matter, radical chemistry, energy materials and more. Muons are created by directing >300 MeV proton beams to a target to produce pions. The pions decay with a mean lifetime of 26 ns via the weak interaction into a muon (or antimuon) and an anti-muon neutrino (or muon neutrino). Due to the parity-violating properties of the weak interaction, all the muons (or antimuons) that are produced are right (or left)-handed, resulting in a spin-polarized muon beam. This spin polarization is important, as it is the gyromagnetic ratio of the muon (135.5 MHz/T) in the material being studied that leads to incredibly sensitive measurement of the local magnetic properties by measuring the Larmor precession of the muon in the material. In practice, most µSR experiments are performed using anti-muons, which decay with a mean lifetime of 2.2 µs, emitting a positron preferentially along the direction of the muon spin at the time of decay. Thus, by measuring the time from implantation to decay of a large number (typically on the order of $10^7$) of muons and the direction of the positron emission, it is possible to reconstruct the time-dependence of the muon spins and thus measure the local magnetic properties.

When performing µSR measurements with pulsed beams, an optimal pulse structure would be as sharp and bright as possible, with sufficient delay between pulses as to allow the muons to decay into positrons. In practice, the muon pulse will be broadened by the pion lifetime, and so a proton pulse on the muon target of the same width (approximately 30 ns) will not cause a loss of resolution. Allowing 20 µs (9 muon lifetimes) between pulses (50 kHz pulse repetition rate) is also sufficient to reduce the background as 99.995% of the muons will have decayed.

Recently, investigations have been made on the feasibility of co-located µSR facility at the Spallation Neutron Source (SNS) at Oak Ridge National Laboratory [1,2]. The high power (1.4 MW) and current (50 mA) of the SNS accelerator would allow for a world-class µSR source to be constructed, however the conventional approach would require a parasitic muon target – placed upstream from the neutron target – reducing the flux of neutrons for the current SNS instrument suite. At existing facilities where both muon and neutron beams are produced, the muon target is

20-30 m upstream from the neutron target and uses 2-3% of the proton beam, with the remaining protons passing through the muon target to be used for neutron production [3]. However, this causes compromises for both neutron and muon beams; the flux on the neutron target is reduced ~20% due to scattering through the muon target, and the proton pulse on the muon target is too long, usually 20-30 µs wide and requiring the muon pulse to be chopped to be usable. We proposed to use laser-based proton extraction to power a muon source at SNS without using a parasitic, upstream target.

Fig. 1 shows a schematic of the proposed muon generation using this method. The proton beam is generated by using laser neutralization of H- beam in the high energy beam transport (HEBT) section of the SNS accelerator beamline. The H- beam energy is 1 GeV in the current operation and will be elevated to 1.3 GeV after completion of the ongoing proton power upgrade project. In addition to being non-intrusive and re-directing only 0.2% of the beam, the muon pulse structure is completely determined by the laser pulse and can be easily controlled by electronics. This novel approach would avoid the conventional issues noted above, while simultaneously not intruding on any of the current missions of the SNS complex.

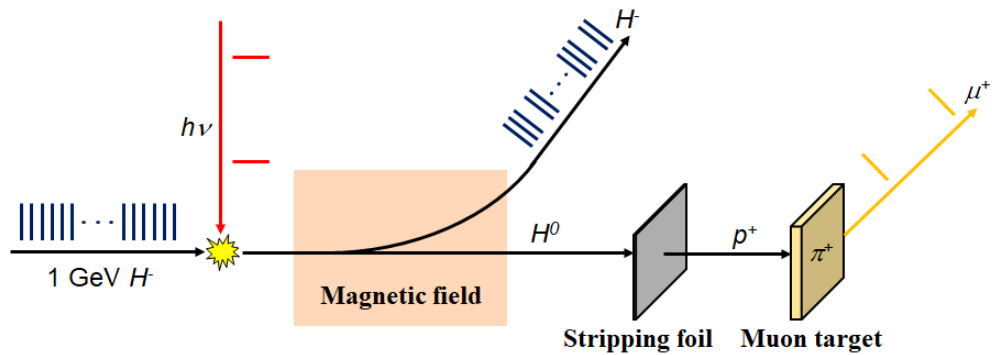

Figure 1: Schematic of laser-based proton pulse extraction from the SNS accelerator beam line for muon generation.

This paper describes a proof-of-principle demonstration of 1 GeV proton pulse extraction from the SNS accelerator beam line based on laser neutralization. A fiber/solid-state hybrid laser system has been developed to generate 30 ns/50 kHz laser pulses with flexible control of the pulse structure and scalability to higher laser power. The laser neutralization experiment has been conducted at the HEBT of the SNS accelerator and the temporal structure of the extracted $H_0$ and proton beam has been verified by measuring detached electrons from laser and metal wire. Discussions on high-efficiency proton pulse generation have been made based on the experimental results.

## 2. Laser System

The non-intrusive proton beam extraction is based on the neutralization of hydrogen ions using a laser. The photo-neutralization yield is proportional to the product of the photon and ion densities in the photon-ion interaction region [4,5] and the neutralization efficiency can be directly calculated by integrating the product between the photon and ion densities over the entire space and time of the laser-ion interaction. Using Eqs. (1)-(4) in [4] and assuming a collimated laser beam size in the interaction region, we calculate the neutralization efficiency as

$$\eta = \frac{\sigma E_l \lambda}{2\pi hc} \cdot \frac{1}{\sqrt{\sigma_{bz}^2 + \sigma_{lz}^2}} \cdot \frac{1}{\sqrt{\sigma_{bx}^2 + \sigma_{lx}^2 + \beta^2 \sigma_{by}^2 + \beta^2 \sigma_{ly}^2}}, \tag{1}$$

where $\sigma$ is the neutralization cross section, $E_l$ is the laser pulse energy, $\lambda$ is the laser wavelength, $h$ is the Planck constant, $c$ is the light speed, $\beta$ is the relativistic factor of the H- beam, and $\sigma_{bx,y,z}$ ($\sigma_{lx,y,z}$) are the RMS beam sizes of the H- (laser) beam in the respective direction. Here we assumed that H- beam is propagating in the *x*-axis direction while laser light is propagating in the *y*-axis direction. Typical H- beam parameters at the HEBT section of the SNS accelerator are listed in Table I. Note that $\sigma \approx 2.9 \times 10^{-17}$ cm2 for $\beta=0.875$ and $\sigma_{ly}$ ($\sigma_{bx}$) are calculated from the pulse duration of laser (H-) beam. A neutralization efficiency above 10-3 would be needed to accurately characterize the temporal structure of the extracted $H_0$ and proton pulses. For a laser with parameters in Table I, a peak power around 100 kW will be required.

The baseline H- beam of the SNS accelerator are 1 ms/60 Hz macropulses and each macropulse consists of 50 ps/402.5 MHz micro pulses. To minimize the laser average power, an ideal laser system needs to operate in a burst mode with an identical macropulse pattern to the H- beam. In each macropulse, laser beam is further bunched into 30 ns/50 kHz micro pulses. Such a macropulse laser can be developed using a master oscillator power amplifier (MOPA) scheme which consists of a seed laser, a fiber-based preamplifier, and solid-state based power amplifiers [6-11]. Pulse generation in a MOPA laser typically consists of two steps: the fast, micro pulses are generated from the seed laser through direct current modulation of a semiconductor laser [10] or external modulation using an electro-optic modulator (EOM) [6,11] while the slow, macropulses are formed by using a pulse picker to select a group of micro pulses. However, when amplifying 30 ns/50 kHz laser pulses using a fiber amplifier, the low-duty-cycle pulses leave gain available in the fiber for long durations between pulses, which can lead to parasitic lasing or destructive self-pulsations that causes damage to the fiber. In [11], auxiliary pulses were added to the seed laser to maintain the duty factor using feedback electronics.

Table I: Parameters of H- beam at HEBT of SNS accelerator and laser parameters proposed in the proof-of-principle demonstration.

| H- beam parameters | | Laser beam parameters | |
|---|---|---|---|
| Beam energy | 1 GeV ($\beta = 0.875$) | Wavelength | 1064.5 nm |
| Beam current | 30 mA | Peak power | Variable |
| Bunch length | 50 ps | Pulse width | 30 ns (FWHM) |
| Repetition rate | 402.5 MHz | Repetition rate | 50 kHz |
| Beam size | $\sigma_{bx} = 18.1$ mm | Beam size | $\sigma_{lx} = 25$ um |
| | $\sigma_{by} = 1.2$ mm | | $\sigma_{ly} = 3.82$ m |
| | $\sigma_{bz} = 0.9$ mm | | $\sigma_{lz} = 25$ um |

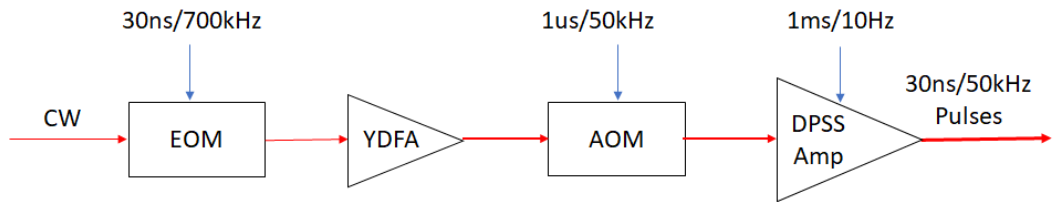

Figure 2: Block diagram of laser system for generation of 30 ns/50 kHz laser pulses.

In this work, we propose to use a combination of electro- and acousto-optic modulators to effectively generate 30 ns/50 kHz laser pulses with uniform pulse shape in both micro and macro waveforms. Fig. 2 shows a block diagram of the proposed laser system. The master oscillator consists of a continuous-wave (CW) fiber laser and an intensity EOM to generate 30 ns pulses at 700 kHz repetition rate. The laser pulses are first amplified by ytterbium-doped fiber amplifiers (YDFAs) and then directed to an acousto-optic modulator (AOM) which selects every 14th laser pulse. The selected 30 ns/50 kHz pulses are amplified by diode-pumped solid-state (DPSS) amplifiers. The initial goal is to achieve 30 ns/50 kHz laser pulses with the peak power exceeding 100 kW. By replacing the current CW laser seeder with a mode-locked laser that provides 50 ps/402.5 MHz pulses, the current laser system design can provide two orders of magnitude higher laser peak power at the same pump laser power.

All RF signals in this setup, i.e., the radio-frequency (RF) control signals of EOM and AOM, as well as the driving currents for both Nd:YAG amplifier modules, are triggered by a multi-channel delay generator. In this way, the phase differences between all RF pulses can be adjusted at a nanosecond accuracy. The precise timing control allows that the 30 ns pulse from the EOM output be selected by the AOM within its stable operation regime and therefore reduces the noise level during the pulse-picking process. The phase control of the Nd:YAG amplifiers also helps to flatten the macropulse waveform. Pulse shaping of individual 30 ns pulses, on the other hand, is implemented by properly controlling the RF pulse waveform applied to the EOM.

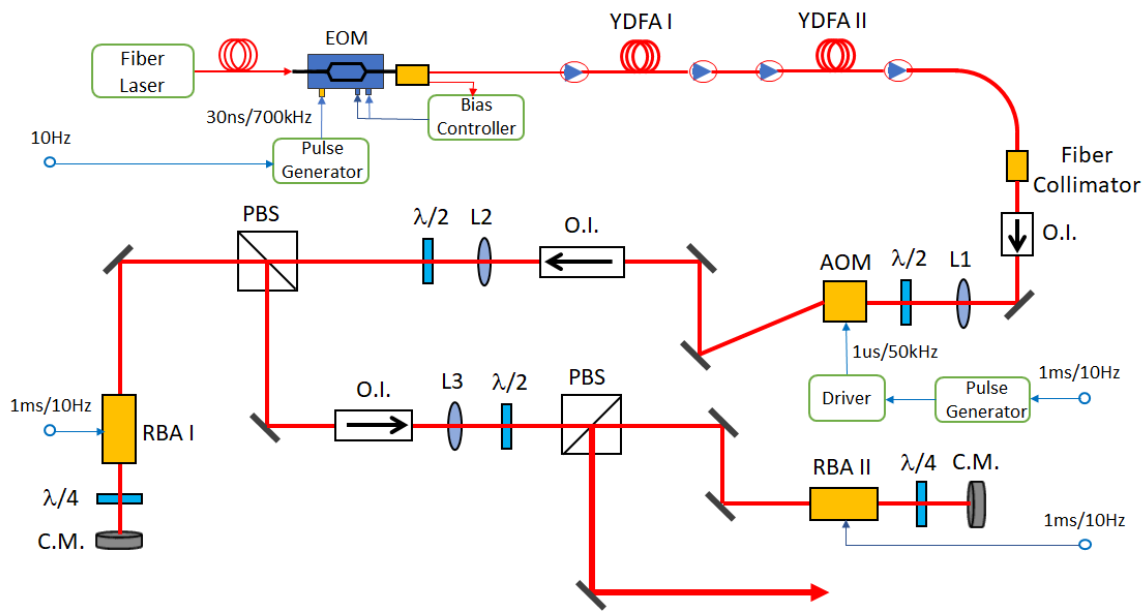

Figure 3: Schematic of laser system. O.I.: optical isolator, L: lens, λ/2: half wave plate, λ/4: quarter wave plate, C.M.: curved mirror, PBS: polarization beam splitter, RBA: diode-pumped Nd;YAG amplifier module.

Fig. 3 shows a detailed scheme of the developed laser system. The front-end laser is a fiber laser (NP Photonics) with the center wavelength at 1064.45 nm and a narrow linewidth (5 kHz). The fiber laser output is modulated by a high-bandwidth EOM (EOSPACE) to produce 30 ns/700 kHz pulses. Two DC bias ports of the EOM are controlled by the YY Labs' dual-base-MZ modulator bias controller [6] which stabilizes the EOM DC voltages using an active feedback circuit. A 40-dB extinction ratio has been achieved with the bias controller. The EOM output has a peak output power of 5 mW and average power of 0.1 mW that is first amplified by two stages of YDFAs. The first YDFA brings the (average) laser power to 10 mW and the second YDFA increases it to 800 mW. No stimulated Brillouin scattering (SBS) oscillations were observed in the YDFAs.

The laser output from fiber amplifiers is coupled into free-space solid-state amplifiers through a fiber collimator. A 30-dB Faraday isolator is positioned immediately after the collimator to eliminate light reflections from down-stream optics to fiber amplifiers. The transmitted laser beam is then re-collimated into a free-space AOM operated at 1 μs/50 kHz. The timing of the AOM driver is properly adjusted so that every 14th pulse of the incoming laser beam is picked. The AOM control signal is further gated to match the macropulse of the SNS H- beam. For the current experiment, the output (diffracted beam) from the AOM has a 30 ns/50 kHz pulses bunched into 1 ms/10 Hz macropulse.

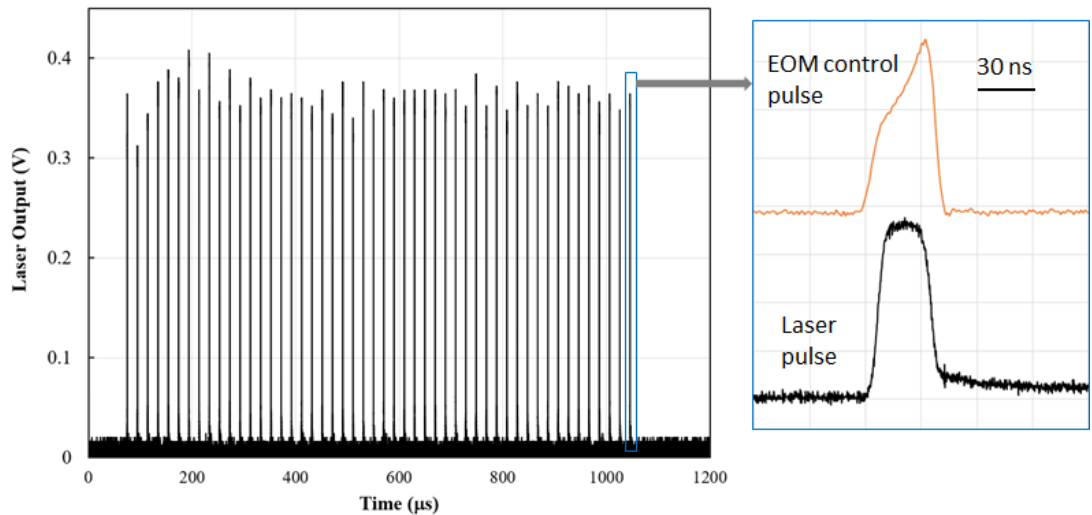

Figure 4: Laser output waveform. Left: 50 pulses generated from the laser; Right: zoom-in of single laser pulse and the corresponding EOM control RF signal.

The 30 ns/50 kHz laser beam from the AOM is amplified by two stages of double-pass diode-pumped Nd:YAG amplifiers (Northrop Grumman RBAT24). The gain medium is a 2 mm-diameter, 4 cm-long Nd:YAG rod. The light after the first pass is propagating through a quarter-wave plate and then reflected by a curved mirror (radius of curvature of 3 m). The reflected light beam, after being rotated 90° in polarization and amplified through the rod, is reflected by the polarization beam splitter (PBS) and redirected to the next stage. Both amplifiers are pumped by diode bars and the driving current is supplied from a programmable driver that provides required pulse width and repetition rate.

Fig. 4 shows a typical waveform of the light output from the final amplifier. A total of 50 pulses are obtained in a 1-ms macropulse. The detailed structure of individual pulse is shown in the inset box together with the waveform of the control signal applied to the EOM. The full width at half maximum (FWHM) of the laser pulse is verified to be 30 ns. The average laser output power is measured to be 3.8 W, corresponding to a peak power of about 250 kW. The spatial profile of the laser beam shows an excellent Gaussian distribution with the beam quality factor $M_2$ measured to be around 1.1.

### 3. H- Neutralization and Proton Generation

Here we describe the initial experiment on neutralization of the H- beam using the developed laser system, extraction of the neutralized hydrogen, or $H_0$ beam, from the H- beam line, and conversion from the $H_0$ beam to protons. The laser beam is sent to the accelerator beam line using an existing free-space laser transport line. The experiment takes place at the HEBT section of the SNS accelerator. Fig. 5 shows the layout of the experiment. The laser-ion interaction chamber is located before the linac dump. Currently, a laser wire-based H- beam emittance measurement station is in this location [12]. The laser beam is focused into a narrow line and interacts with the H- beam at a 90° angle. As a result of the neutralization, electrons of a narrow slice of the H- beam are detached. After the laser-ion interaction chamber, the particle beam (a mix of H- and $H_0$) passes through a dipole magnet. The majority of the beam is the undetached H- ions that propagates along the original orbit. The $H_0$ beam, on the other hand, receives no influence from the magnetic field and follows a straight line into the linac dump. In the linac dump, a wire scanner is installed about 10 meters down-stream of the laser-ion interaction chamber and the titanium wire detaches the electrons from the $H_0$ beam and converts neutral hydrogen atoms to protons. Note that only a portion of the $H_0$ beam is converted to protons by the wire scanner for the purpose of pulse structure verification. A complete $H_0$ to proton conversion can be realized with a stripping foil as is routinely achieved at the injection of the current SNS accumulator ring [13].

The synchronization of laser pulses to the SNS accelerator timing card allows an effective detection of the electron signals from the laser neutralization. Two types of signal detection have been employed in the present experiment. The electrons detached from the H- beam by the laser light are detected by the Faraday cup (FC) and the signal is amplified by two stages of amplifiers with a combined gain of about 15 and a bandwidth of close to 100 MHz. The FC output not only measures the temporal profile of the H0 beam, but also provides an estimation of the electron charges as a result of the photo-detachment. A drawback of the FC detector is its high noise floor. The electrons detached by the wire scanner are detected using a combination of scintillator and photomultiplier tube (PMT). This detection scheme has an extremely high sensitivity and therefore is particularly useful in the detection of signals created by weak laser power. Since the PMT/scintillator detection scheme only detects a portion of the detached electrons, the PMT output is used to confirm the temporal structure of the proton beam in the low laser power regime.

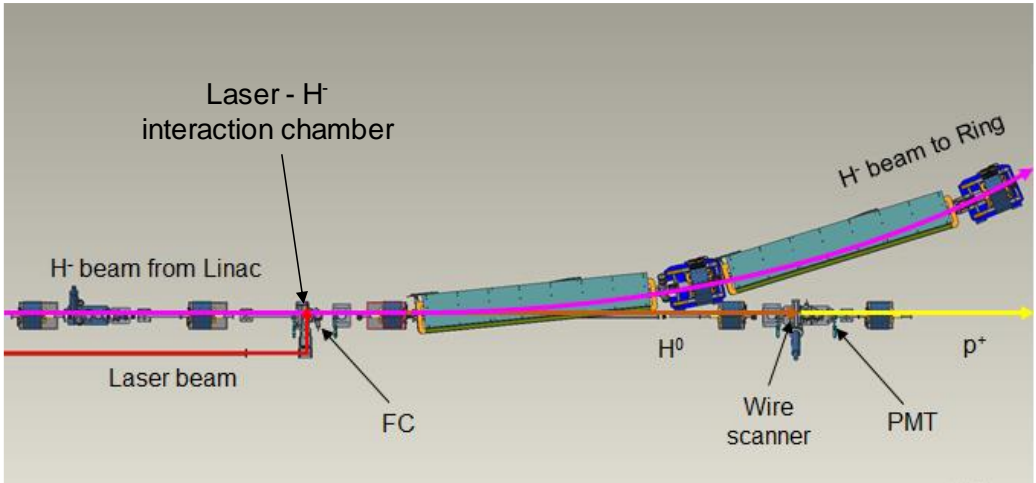

Figure 5: Location of laser-ion interaction chamber. FC: Faraday cup, PMT: photomultiplier tube.

Fig. 6 shows a typical experimental result of the FC output together with the corresponding laser output. As shown in Fig. 6, the positions of the detected electron pulses agree very well with the laser pulses. We noticed that the first electron pulse in the FC output is quite weak due to the lower current at the beginning of the H- pulse ramp up. The fluctuation in the pulse amplitude is possibly related to the position jitter of the laser pulse or the current variation of the ion beam. Details of the individual pulse are shown in the inset box. The results clearly revealed the 30 ns/50 kHz pulse structure.

An example of the PMT output waveform is shown in Fig. 7 for a low laser peak power. The PMT output gives a clear indication of the 30 ns/50 kHz pulse structure of the detected electrons. We note that the pulse broadening shown in the inset box of Fig. 7 is mostly caused by the slow response of the scintillator.

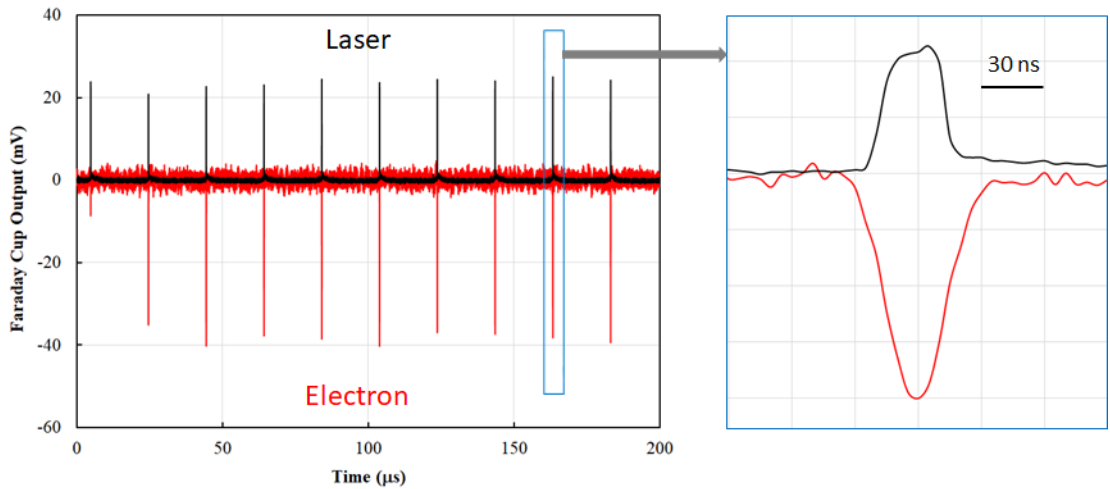

Figure 6: Detected electron pulses by Faraday cup and the corresponding laser pulses. The laser peak power at the interaction point is estimated to be 230 kW.

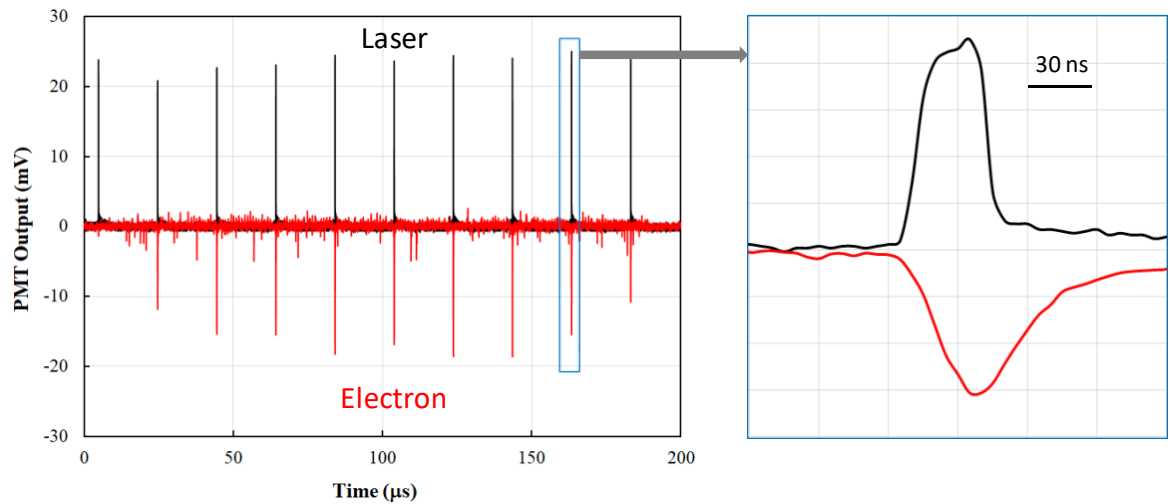

Figure 7: Detected electron pulses by PMT and the corresponding laser pulses. The laser peak power at the interaction point is estimated to be 27 kW.

We have varied the laser power and measured the electron pulse amplitude from both FC and PMT outputs, with the results shown in Fig. 8. We observed the expected linear dependence from both FC and PMT output measurements. Due to the presence of background noise in the FC output, no valid pulse structure can be identified at the laser power below 50 kW. On the other hand, the scintillator/PMT detection shows excellent sensitivity to the signals induced by laser pulses with the peak power as low as 10 kW.

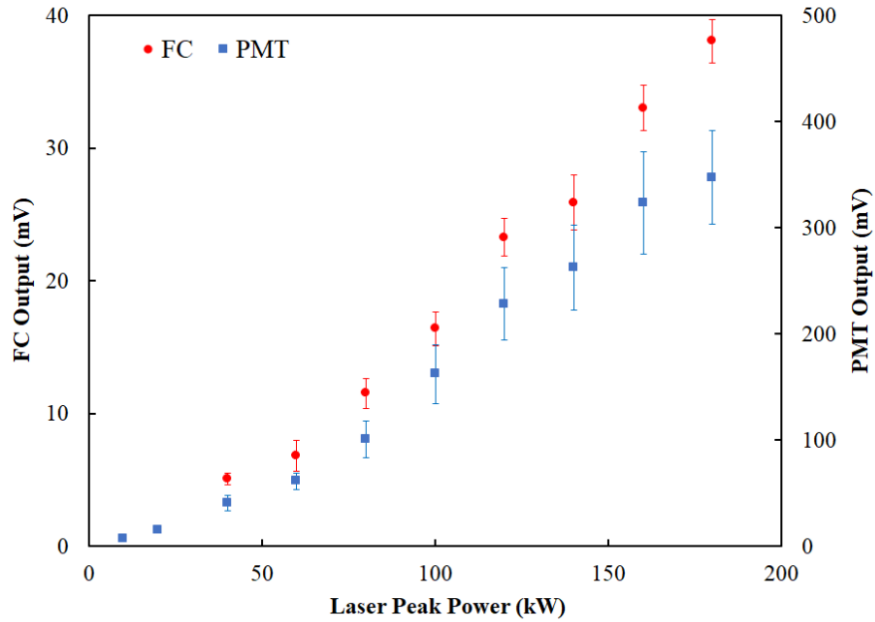

Figure 8: Detected electron pulse magnitude from FC and PMT versus laser peak power. The error bars are the standard deviation of 10 pulses.

4. **Discussions**

We have calculated the photo-detached electron charge from the FC output on a 1 GeV H- beam. The number is compared with the charge of the H- beam within the time window (~30 ns) of the laser pulse and the ratio gives an estimation of the neutralization efficiency obtained in the experiment. For an estimated laser peak power of 80 kW at the laser-ion interaction point, we measured the neutralization efficiency to be ~0.14% which is about 66% of the theoretical prediction from Eq. (1) based on the beam/laser parameters. We consider the difference is possibly due to the non-perfect electron collection by the FC.

To achieve a neutralization efficiency around 90% and hence enable high-flux muon generation based on the extracted proton pulses, a number of improvements will be implemented on the laser system and laser-ion interaction scheme. The first improvement is to increase the laser power by introducing micro-bunch structure to the laser pulse and adding more amplifiers to the current laser system, which can increase the laser peak power to several tens of MW using the current laser configuration. The second improvement is to optimize the laser and H- beam parameters. In particular, the vertical beam size match between the laser and H- beams is important in achieving high neutralization efficiency while maintaining moderate laser intensity on optics surfaces. Another improvement is the recycling of the laser power by using multi-reflections of laser beam in a non-resonant optical cavity [11]. Due to the low cross-section value in the neutralization, laser power recycling is both necessary and highly feasible.

A diagram of the temporal structure matching between laser and neutron production H- beam at SNS is shown in Fig. 9. For high-efficiency laser neutralization of 1.3 GeV H- beam, we assume that the H- beam is compressed in the vertical direction to half of its current value and the laser beam is collimated to match the vertical beam size of the ion beam, i.e., $\sigma_{lx} = \sigma_{lz} = \sigma_{bz} = 0.5$ mm, $\sigma_{ly} = 25.5$ mm. The rest parameters are the same as in Table I. Numerical calculations indicate that a 23.7% neutralization efficiency can be achieved in a single laser-ion interaction for a laser peak power of 10 MW. The average laser power in this case is 77 W when the laser is operated at the full cycle of the SNS operation, i.e., 1 ms/60 Hz. Such a laser power can be readily achieved from the proposed MOPA laser system using diode-pumped Nd:YAG amplifiers. Meanwhile, the resulting laser intensity in this case is only 0.08 J/cm$_2$ which is well below the damage threshold of the standard optics. If a non-resonant multi-reflection optical cavity is employed, only 10 reflections would be needed to bring the neutralization efficiency to over 93%.

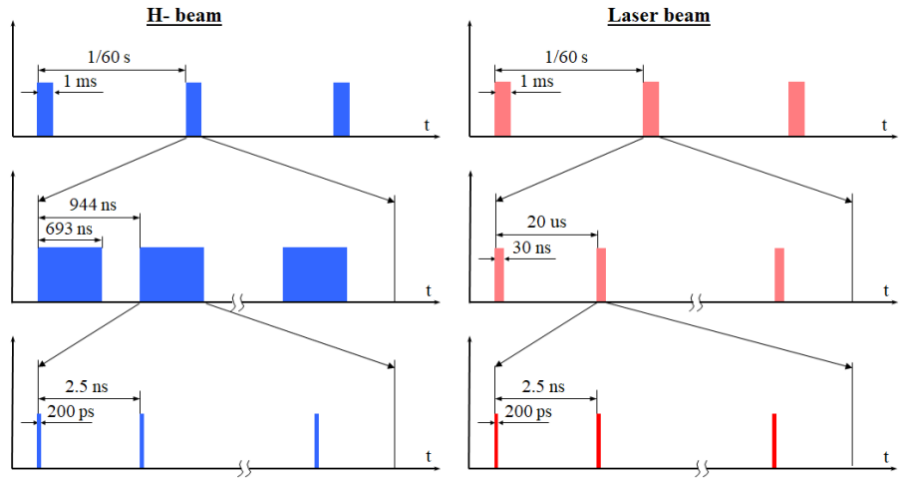

Figure 9: Temporal structure of neutron production H- beam at SNS (left column) and proposed laser pulse (right column).

Furthermore, using the proton pulse parameters and efficiency that were demonstrated in this work allows for an estimate of the muon flux that could be achieved with this facility. Previously, Monte Carlo simulations have been carried out to determine an optimal target material and geometry for muon production utilizing 1.3 GeV protons that will be present after the ongoing proton power upgrade of the SNS accelerator [14]. The simulations predict a surface muon conversion rate of $1.3 \times 10^{-4}$ $\mu_+/p_+$, comparable to rates calculated for other muon targets [15]. This means that surface muon fluxes from an optimized target could exceed $10^9$ $\mu_+$/sec (at a 50% neutralization efficiency), which is orders of magnitude higher than any currently existing pulsed µSR facility. Coupled with the high resolution that would be enabled by utilizing the laser extraction, a pulsed muon facility at the SNS that utilizes this method of proton pulse generation would be a world-leading facility for µSR experiments.

## 5. Conclusion

We have described a non-intrusive method of high-energy proton pulse extraction from the SNS accelerator beam line by using laser neutralization of H- beam. Proton pulses could be used to generate a pulsed muon beam with negligible influence on the present neutron production at SNS. A muon facility based around this system would utilize only 0.2% of the SNS proton beam, a stark contrast from the 20% reduction seen at other co-located muon and neutron facilities. The temporal structure of the proton beam can be controlled by the laser and can be matched to optimize the performance for µSR applications. A laser system based on a fiber-based master laser oscillator and diode-pumped solid-state laser amplifiers has been developed to generate 30 ns/50 kHz laser pulses. A laser neutralization experiment has been implemented and the proton pulses with the temporal structure identical to the laser beam has been verified at the SNS linac dump. The scalability to over 90% neutralization efficiency based on the current method has been discussed. Our experiment strongly supports the feasibility of a co-located µSR facility at the SNS and predicts that a facility based on the laser neutralization design would be the world's highest resolution and highest flux pulsed µSR source.

**Acknowledgements**

The authors acknowledge A. Webster, S. Murray III, A. Dawson for their help and D. Johnson for useful discussions. We are grateful for the support by A. Aleksandrov, S. Cousineau, and F. Pilat. This work was supported in part by the ORNL LDRD Seed program. ORNL is managed by UT-Battelle, LLC, under contract DE-AC05-00OR22725 for the U.S. Department of Energy.